# Time-stretch infrared spectroscopy

Akira Kawai,[1] Kazuki Hashimoto,[2] Venkata Ramaiah Badarla,[2] Takayuki Imamura,[1] and Takuro Ideguchi,[1,2,3,*]

[1]Department of Physics, The University of Tokyo, Tokyo, Japan
[2]Institute for Photon Science and Technology, The University of Tokyo, Tokyo, Japan
[3]PRESTO, Japan Science and Technology Agency, Saitama, Japan
[*]Corresponding author: ideguchi@ipst.s.u-tokyo.ac.jp

**Improving spectral acquisition rate of broadband mid-infrared spectroscopy promises further advancements of molecular science and technology. Unlike the pump-probe spectroscopy that requires repeated measurements with different pump-probe delays, continuous spectroscopy running at a high spectral acquisition rate enables transient measurements of rapidly changing non-repeating phenomena or statistical analysis of a large amount of spectral data acquired within a short time. Recently, Fourier-transform infrared spectrometers (FT-IR) with rapid delay scan mechanisms including dual-comb spectrometers have significantly improved the measurement rate up to ~1 MSpectra/s that is fundamentally limited by the signal-to-noise ratio. Here, we overcome the limit and demonstrate the fastest continuous broadband vibrational spectrometer running at 80 MSpectra/s by implementing wavelength-swept time-stretch spectroscopy technique in the mid-infrared region. Our proof-of-concept experiment of the time-stretch infrared spectroscopy (TS-IR) demonstrates broadband absorption spectroscopy of phenylacetylene from 4.4 to 4.9 μm (2040-2270 cm$^{-1}$) at a resolution of 15 nm (7.7 cm$^{-1}$) with a superior signal-to-noise ratio of 85 without averaging and a shot-to-shot fluctuation of 1.3%.**

Vibrational molecular spectroscopy has been used as an indispensable tool for investigating molecular systems in a variety of fields. The pump-probe vibrational spectroscopy has brought tremendous advancements in physical and biological chemistry with the femtoseconds to picoseconds temporal resolution, but it can be only applicable to repetitively reproducible phenomena with the pump pulse[1-4]. Measuring rapidly changing dynamics including stochastic or non-repeating phenomena such as gaseous combustion or conformation change of proteins necessitates high-speed continuous measurement techniques. High-speed continuous measurement is also essential for analyzing a large number of spectra with, for example, raster scan imaging[5] or flow cytometry[6,7]. Among various vibrational spectroscopy techniques, single detector based methods have gained an advantage in measurement speed because a fast photodetector with the bandwidth of ~MHz-GHz outperforms the scan rate of a grating based dispersive spectrometer with a line sensor operated at a rate of ~kHz. In the last couple of decades, we have especially witnessed significant advancements of broadband coherent Raman scattering spectroscopy, but its scan speed is essentially limited by the low SNR, resulting in ~1-10 kHz measurement rates[5,8-14].

Mid-infrared (MIR) absorption spectroscopy, the counterpart modality of vibrational spectroscopy to Raman spectroscopy, holds much potential for the high-speed continuous measurement because of its inherently larger cross-section, several orders of magnitude larger than that of the coherent Raman scattering. In the MIR region, where the richest molecular signatures exist, Fourier-transform spectroscopy techniques with rapid delay scan mechanisms including dual-comb spectroscopy[15,16] have been the fastest methods to date[17,18]. To the best of our knowledge, the fastest measurements with a sufficient spectral resolution of ~100 GHz for measuring liquid phase molecules and a signal-to-noise ratio (SNR) of ~10-100 were recorded by dual-comb spectroscopy with quantum cascade laser (QCL) combs or microresonator combs, which are operated at acquisition rates of ~MSpectra/s[19,20]. Importantly, the acquisition rates of these systems are not limited by their instrumental scan rates but SNR[21,22], which fundamentally prevents further improvement. Therefore, to improve the measurement speed, one must use another technique that is able to provide a better SNR.

Wavelength-swept spectroscopy is known to have an advantage in SNR by a factor proportional to $\sqrt{N}$, where $N$ is the number of spectral elements, against the multiplex measurement such as Fourier-transform spectroscopy[23]. Therefore, especially for broadband spectroscopy with a large $N$, the SNR advantage becomes significant. In the MIR region, wavelength-swept lasers such as external-cavity QCLs have been demonstrated, but these lasers have shown limited instrumental scan rates up to 250 kHz[24,25], which provide lower spectral acquisition rates than those of the state-of-the-art dual-comb spectrometers. Now, we recognize that a single-pulse broadband spectroscopy technique called time-stretch spectroscopy (also known as dispersive Fourier-transform spectroscopy) is an ideal frequency-swept spectroscopy technique that can be operated at a rate of 10s MSpectra/s with a femtosecond mode-locked laser[26]. However, this technique has been demonstrated only in the near-infrared region where advanced telecommunication optics such as extremely low-loss optical fibers and ultrafast photodetectors are available. To the best of our knowledge, it has never been demonstrated in the MIR region due to the lack of suitable technology. To demonstrate MIR time-stretch spectroscopy, there are three components which must be prepared: (1) a high repetition rate MIR femtosecond laser source running at 10s MHz, (2) a low loss time-stretch dispersive optics in the MIR, and (3) a fast MIR photodetector at a bandwidth of several GHz. Especially, the last two components are technically demanding. For example, an optical fiber, which is usually used as a time-stretcher, is too lossy in the MIR (~100-1000 dB/km), and bandwidths of commercially available fast MIR photodetectors are too low (~100 MHz-1 GHz). Note that the fast photodetector is not necessarily required for demonstrating the higher SNR of TS-IR but needed for achieving the high-speed measurement capability.

In this work, we demonstrate time-stretch infrared spectroscopy (TS-IR) spanning over 4.4-4.9 μm running at the fastest acquisition rate of 80 MSamples/s, 1-2 orders higher than the previous record enabled by the inherently better SNR of the wavelength-swept spectroscopy. In our system, we use (1) a synchronously-pumped femtosecond optical parametric oscillator (fs-OPO) as a MIR light source, (2) a pulse-stretching technique termed free-space angular-chirp-enhanced delay (FACED)[27,28], and (3) a recently developed quantum cascade detector (QCD) with an extremely high -3-dB bandwidth of 5 GHz[29].

Our TS-IR spectrometer is schematically shown in Fig. 1. Our light source is a home-made MgO:PPLN fs-OPO pumped with a Ti:Sapphire mode-locked laser running at a repetition rate of 80 MHz (Maitai, Spectra Physics). The OPO cavity is designed for resonating the signal pulses. Wavelength of the MIR idler pulses can be tuned from 2.1 to 5.1 μm by changing the OPO cavity length, the PPLN grating period and/or the pump wavelength. In this work, we set the center wavelength of the idler pulses at 4.6 μm. An average power of the MIR pulses is 70 mW. The pulse stretching is made by FACED system, which consists of a diffraction grating (219 grooves/mm, Richardson Grating), a pair of concave mirrors (f=100, and 200 mm) and a pair of flat mirrors with a separation distance of $d$ (mm) and an angle of α (degree). All the above optics are coated with gold. In FACED system, wavelength dependent incident angle of the beam into the flat mirror pair is translated into the chromatic optical path length delay (linear chirp), leading to the pulse stretch from ~100 fs to ~10 ns along with the wavelength-time conversion. The output pulse consists of sub-pulses due to a discrete condition of back reflection at the flat mirror pair. The power throughput of the FACED system is about 9%. This high throughput is especially important in the case of MIR pulse stretching because a conventional technique using a single-mode fiber suffers from significantly large optical loss. The stretched pulses with the average power of 6.3 mW are picked out with a beamsplitter and focused onto a photodetector. We use an uncooled QCD with a -3-dB cutoff frequency of 5 GHz (20 GHz for -20 dB) made by Hamamatsu Photonics K.K. as a fast MIR photodetector. The QCD can be operated with much larger bandwidth than the conventional semiconductor based detectors due to the short lifetime of the sub-band transitions. Also, the non-biased operation leads to the low noise property. Our detector has sensitivity from 3.3 to 6.0 μm. The QCD signal is amplified by 20 dB with a microwave amplifier with a bandwidth of 0.01-26.5 GHz (Hewlett Packard) and digitized at a sampling rate of 80 GS/s by a high-speed oscilloscope with a bandwidth of 16 GHz (Teledyne LeCroy).

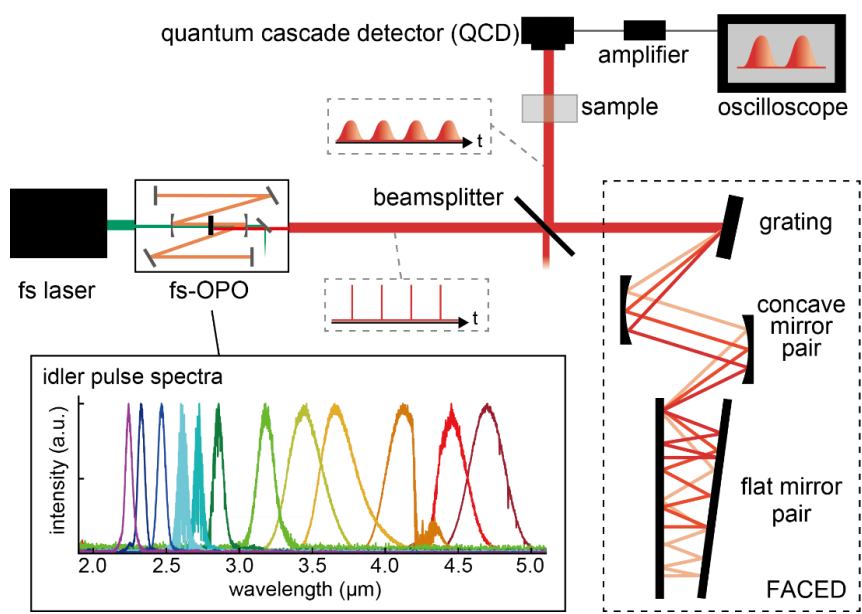

**Fig. 1. Schematic of TS-IR spectrometer.** The MIR fs pulse is stretched in the FACED system and detected by the QCD after passing through the sample. The QCD signal is amplified and digitized with the fast oscilloscope. The bottom left inset shows the wavelength-tunable spectra of the idler pulses generated by the fs-OPO.

To characterize the system, we first measure the stretched pulse waveform under several conditions of a pair of flat mirrors in our FACED system (Fig. 2). We confirm that change in temporal interval between the adjacent sub-pulses ($\Delta t$) depends on the distance between the mirrors ($d$), which agrees well to the theoretical relation

$$\Delta t \sim 2d/c,$$

where c is the speed of light[28]. We also confirm change in number of sub-pulses (spectral components) ($N$) depends on the angle of the mirrors ($\alpha$)

$$N \sim \theta/\alpha,$$

where $\theta$ is a constant value determined by a FACED geometry. In the following experiment, we align the mirror pairs to fulfill the condition, $\Delta t$=0.26 ns and $N$=30.

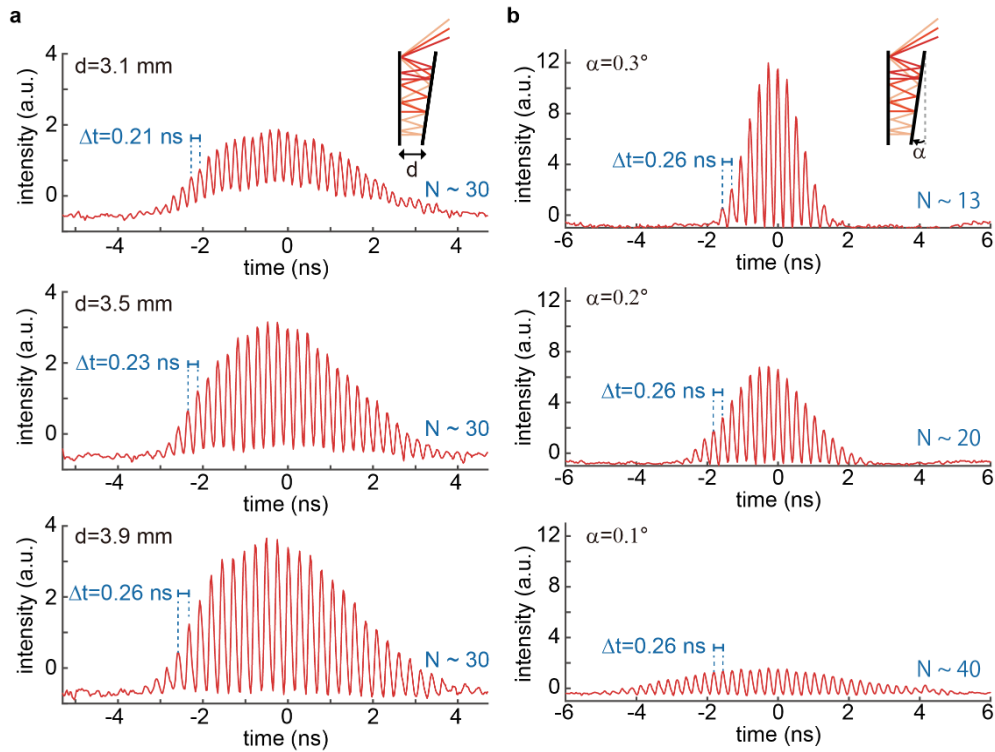

**Fig. 2. Characterization of a time-stretched MIR pulse. a** Mirror-pair-distance ($d$) dependence on the temporal interval of the sub-pulses ($\Delta t$). **b** Mirror-pair-angle ($\alpha$) dependence on the number of sub-pulses ($N$).

To demonstrate spectroscopic capability of TS-IR, we first measure notch-filtered spectra by putting a sharp 0.3-mm beam block near the Fourier plane of the FACED system (Fig. 3(a)). We confirm that a dip in the waveform shifts by 13 nm in wavelength by shifting the position of the beam block by 0.5 mm in the Fourier plane, which agrees well to the theoretical estimation. Then, we measure absorption spectrum of liquid phenylacetylene. The repetitive broadband time-stretched spectra are clearly observed every 12.5 ns (repetition rate of 80 MHz) as shown in Fig. 3(b). The magnified single spectrum (Fig. 3(c)) shows an absorption line characteristic to the C-C triple-bond of the molecules, which agrees well to the reference spectrum measured by a conventional Fourier-transform infrared (FT-IR) spectrometer (Fig. 3(d)). The spectral resolution is 15 nm (7.7 cm$^{-1}$), which is determined by spectral gap between the adjacent spectral elements (an experimental verification is shown in Figure 4), while SNR of the spectrum is evaluated as 85. The shot-to-shot fluctuation of the spectra is evaluated as 1.3% by taking standard deviation of peak intensities of the consecutively measured 40 spectra. The dominant noise source is attributed to the detector and relative intensity noise for lower and higher optical power in our experimental condition, respectively.

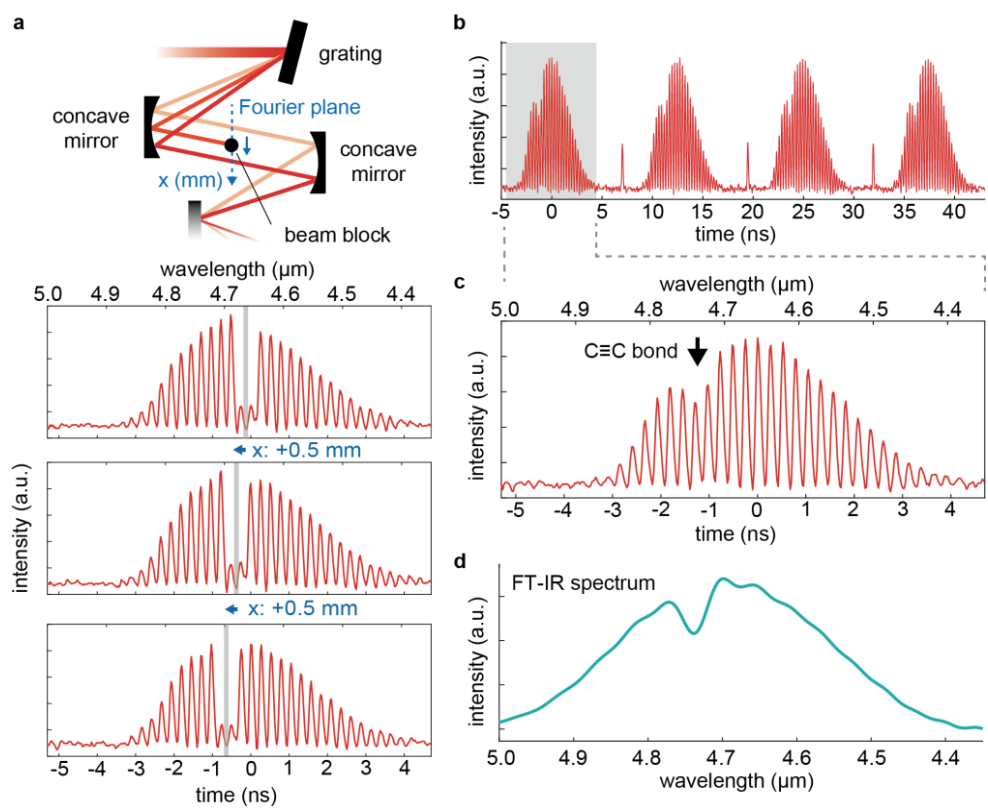

**Fig. 3. TS-IR spectroscopy. a** Spectrally filtered TS-IR spectra. **b** Continuously measured molecular absorption TS-IR spectra of phenylacetylene. **c** A magnified spectrum of (b). **d** A reference absorption spectrum of phenylacetylene measured by FT-IR.

To carefully evaluate the spectroscopic capability of TS-IR, we perform additional demonstrations. We first measure multiple absorption lines from different molecular species, a mixture of phenylacetylene and toluene in a 1-mm thick cuvette with a concentration ratio of 3:47 within the spectral span of the spectrometer (Fig. 4a). A fundamental absorption band of phenylacetylene and an overtone band of toluene are clearly observed, showing the capability of multi-species detection. Next, we evaluate the spectral resolution by measuring a notch-filtered spectrum with narrow linewidth of 13 nm (6.0 cm$^{-1}$), which is separately evaluated by a FTIR measurement at a resolution of 10 nm (4.5 cm$^{-1}$), prepared by sharp beam blocks placed near the Fourier-plane. The TS-IR spectrum shows two sharp lines (dips) with single spectral points, verifying the resolution is determined by the space between the adjacent spectral points, 15 nm (7.7 cm$^{-1}$) as theoretically expected (Fig. 4b). Finally, we verify the linear dependence of the absorbance on the molecular concentration by measuring phenylacetylene diluted by toluene in a 100-μm thick cuvette with different concentrations. Absorbance of a vibrational band of the phenylacetylene is plotted for each concentration in Fig. 4c, showing a linear dependence as expected in linear absorption spectroscopy.

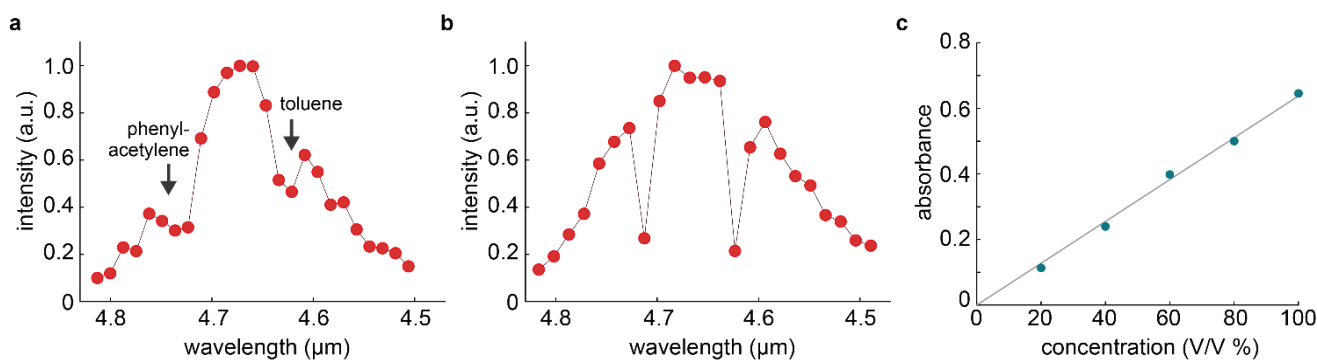

**Fig. 4. Performance evaluation of TS-IR spectroscopy. a** A non-averaged TS-IR spectrum of two molecular species, toluene and phenylacetylene. **b** A non-averaged TS-IR spectrum of narrow notch filters. The observed dips with single spectral points indicate our TS-IR system has a spectral resolution determined by the spectral space between the adjacent spectral point, 15 nm (7.7 cm$^{-1}$) in this experiment. **c** Linear concentration dependence of absorbance evaluated with phenylacetylene diluted in toluene. For the measurements of **a** and **b**, the average power of the OPO output was 55 mW.

The spectroscopic specification of our system can be improved in several directions. First, using a higher repetition rate laser can increase the spectral acquisition rate. Second, since the spectral resolution is currently limited by the detector bandwidth (5 GHz in our case), using a faster detector improves them without sacrificing the measurement speed. Also, simply reducing the repetition rate of the laser leads to the resolution improvement without replacing the detector. Third, the spectral bandwidth can be enlarged by using a laser with a broader spectrum and multiple QCDs that cover the broad spectrum.

Potential applications of our ultra-rapid and broadband MIR spectrometer are versatile. For example, in the field of biochemical study on photoreceptive proteins such as rhodopsin where the dynamics of molecular conformation change is studied, the nanosecond to microsecond is a missing time-scale especially for measuring non-repeatable one-way structural change[2,30]. The TS-IR could open a door to this unexploited field with its continuous high-speed measurement capability. In addition, label-free high-throughput measurements such as liquid biopsy of human blood[31] or flow-cytometry for single cell analysis and sorting[7,32] are also promising directions for future biomedicine because TS-IR's continuous high-speed capability allows us to measure a large number of events within a unit time. Furthermore, the TS-IR system can demonstrate an extremely rapid-scan MIR OCT where a larger penetration depth is available than the conventional OCT systems[33].

**Data availability**

The data provided in the manuscript are available from T.I. upon request.

**Acknowledgement**

The authors thank Makoto Kuwata-Gonokami, Junji Yumoto and Yu Nagashima for letting us use their equipment. This work was financially supported by JST PRESTO (JPMJPR17G2), JSPS KAKENHI (17H04852, 17K19071), Research Foundation for Opto-Science and Technology, and Murata Science Foundation.